\newcommand{\be}{\begin{equation}}
\newcommand{\ee}{\end{equation}}
\newcommand{\bea}{\begin{eqnarray}}
\newcommand{\eea}{\end{eqnarray}}
\newcommand{\bra}[1]{\big< #1 \big|}
\newcommand{\ket}[1]{\big| #1 \big>}
\newcommand{\braket}[2]{\big< #1 \big| #2 \big>}
\newcommand{\rbra}[1]{\big( #1 \big|}
\newcommand{\rket}[1]{\big| #1 \big)}
\newcommand{\Pslash}{ P \mskip -11mu / }
\newcommand{\qslash}{ q \mskip -8mu / }
\newcommand{\nn}{\nonumber}
\documentstyle[11pt]{article}
\title{OPE analysis for polarized deep inelastic scattering}
\author{B.~Ehrnsperger$^1$, L.~Mankiewicz$^{1,2}$, A.~Sch\"afer$^1$
\\
$^1$Institut f\"ur Theoretische Physik, J.~W.~Goethe
Universit\"at Frankfurt,
\\
Postfach~11~19~32, W-60054~Frankfurt am Main, Germany
\\
$^2$N. Copernicus Astronomical Center, Bartycka 18, PL--00--716 Warsaw,
Poland}
\begin{document}
\maketitle
\begin{abstract}
We present an explicit OPE analysis for the first moment of $g_1$
up to order $M^2/Q^2$.
This result allows to calculate power corrections to the
Bjorken  and Ellis--Jaffe sum rules.
\end{abstract}
\\ \\
The investigation of the spin structure of the nucleon is one of the
most active fields in hadron physics.
This is demonstrated by the fact that CERN, SLAC, and DESY
all have programs in polarized lepton--hadron scattering.
So far data are available from EMC (proton) \cite{EMC}, SMC (deuteron)
\cite{SMC}, and E142 (helium) \cite{E142}.
As all of these groups
took data at different values of $Q^2$ these results can not be
compared directly.
In particular for E142 the lowest $x$ points are measured at
only $Q^2 \approx 1 \enspace {\rm GeV}^2$ such that higher twist
corrections should be important. Actually one has to face a typical
track--off. At low $Q^2$ one can get very good statistics but the
theoretical analysis is more involved, at high $Q^2$ theory is easy,
but the statistics of the experiment does not allow detailed
statements.

Polarized scattering has, however, the great advantage that the higher
twist corrections seem to be calculable with a sufficient precission
\cite{IIBa90}.
As also the HERMES experiment at HERA will run at small
$Q^2$ it is a necessary task for theory to clarify the issue of $Q^2$
dependence.
The most basic block for this is the operator product
expansion for the moments of $g_1$ and $g_2$.
Unfortunately there was a mistake in the original work
\cite{ESh82} as noticed by Ji and Unrau \cite{XJi93}.
The whole discussion was plagued by the fact
that often the relevant definitions and conventions are not stated
explicitely.
Unluckily the real or apparent differences left often the impression
that the higher twist corrections are kind of undefined.
This is not true.
OPE gives a unique result and by now all groups working on this
problem have reached the same results.
In this paper we present our calculation  in detail with all definitions
given, such that no misunderstandings should remain.

As indicated the aim of this paper is to identify higher twist
operators, which contribute at the next--to leading twist level.
We do not however attempt to calculate $O(\alpha_s )$ corrections to
the Wilson coefficients.

The starting point of the calculation is the forward
virtual Compton scattering amplitude:
\be \label{eq1}
T_{ \mu \nu } (q, P, S) =
{\rm i} \int d^4 \xi \; e^{{\rm i} q \cdot \xi}
\bra{PS} T \left( j_\mu (\xi) \; j_\nu (0) \right) \ket{PS} \; .
\ee
The target nucleon has four momentum $P^\mu$ ($P^2 = M^2$ ) and spin $S^\mu$
($S^2 = - M^2 \; , \enspace S \cdot P = 0 $).
The virtual photon has four momentum $q^\mu$ , ($q^2 = - Q^2 \leq
0$).
The nucleon state is normalized covariantly $\braket{PS}{P' S'} =
2P^0 (2\pi^2) \delta^3 (\vec{P} -\vec{P}') \delta_{SS'} $.
Using the
analyticity properties of $T_{ \mu \nu } $
\be \label{eq2}
\frac{1}{ 2 { \rm i } }
\left[ T_{ \mu \nu }
(q_0 +
{ \rm i }
\varepsilon )
- T_{\mu \nu}
(q_0 - {\rm i} \varepsilon )
\right]
 = - 2 \pi W_{\mu \nu}
\ee
and equation (\ref{eq1}) we find the following form for the hadronic
tensor:
\be \label{eq3}
W_{\mu \nu} = \frac{1}{4 \pi} \int d^4 \xi \;
e^{{\rm i} q \cdot \xi }
\bra{PS}  \left[ j_\mu (\xi) , j_\nu (0) \right]_- \ket{PS}  \; .
\ee
We define commutators and anticommutators as $ \{ A , B \}_+ = AB + BA $
and $ [ A , B]_- = AB - BA $.
The antisymmetric part of the hadronic tensor can be decomposed as
follows:
\be \label{eq4}
W^A_{\mu \nu} = {\rm i} \varepsilon_{\mu \nu \sigma \lambda}
\frac{q^\lambda}{\nu} \left( (g_1(x,Q^2) + g_2 (x,Q^2) ) S^\sigma -
g_2(x,Q^2) P^\sigma \frac{q \cdot S}{\nu} \right)  \; ,
\ee
with $\nu = p \cdot q$,
the Bjorken variable $x = Q^2 /2\nu$ and $ \varepsilon^{0123} = 1$.

We are only interested in $g_1$ and $g_2$, so we
consider only the antisymmetric part of $T_{\mu \nu}$.

Performing a Cauchy integration one obtains the connection between
$T^A_{\mu \nu}$ and $W^A_{\mu \nu}$, which scematicaly reads
\cite{RJa90}:
\be  \label{eq10a}
\int_0^1 W^A_{\mu \nu} x^n \; dx = \frac{1}{8 \pi {\rm i}} \oint
d \omega \; T^A_{\mu \nu}  (\omega) \omega^{-n -2} \; ,
\ee
with $\omega = 1/x$.
Equation (\ref{eq10a}) holds for $n = 0$ only if $T^A_{\mu \nu}
(\omega)$ vanishes fast enough for $\omega \to \infty$.
General considerations based on Regge theory suggest that the part of
$T^A_{\mu \nu}$ that determines $g_1$ goes rapidly enough to zero, but
the part, which determines $g_2$ may not \cite{RJa90}.
In this publication we assume the validity of equation (\ref{eq10a})
for all $n \in {\bf N}_0 $.

Starting from equation (\ref{eq1}) one obtains a representation of
$T^A_{\mu \nu}$ as a series in $\omega$ in the kinematical domain
where $\mid \omega \mid \leq 1$,
i.e.:
\be \label{eq11}
T^A_{\mu \nu} = \sum_m b_{\mu \nu }^{(m)} \omega^m \; .
\ee
Where from equation (\ref{eq10a}) it follows that:
\be \label{eq12}
\int_0^1 W^A_{\mu \nu} x^n \; dx = \frac{1}{4} b_{\mu \nu }^{(n+1)} \; .
\ee
The only Feynman amplitudes which contribute to $W^A_{\mu \nu}$ up to
$1/Q^2$ order are shown in Figure 1 \cite{ESh82}.
To proceed further we follow the method proposed by Shuryak and
Vainsthein \cite{ESh82}.
In their formalism contribution of Figure 1 to equation (\ref{eq1})
has the form:
\be \label{eq7}
T^A_{\mu \nu} (q, P, S) = - A_{(\mu \nu)} \sum_f \int d^4 \xi \;
\bra{PS} \rbra{\xi} \bar{\psi}_f q^2_f \left( \gamma_\mu \frac{1}{\Pslash +
\qslash} \gamma_\nu  + \gamma_\nu \frac{1}{\Pslash - \qslash} \gamma_\mu
\right) \psi_f \rket{0} \ket{PS} \; .
\ee
The symbol $A_{(\mu \nu)}$ denotes antisymmetrization of the indices
$\mu $ and $\nu $.
Through this paper we use the Schwinger notation
\cite{JSc51}, in which
$\rket{\xi}$ denotes a formal eigenvektor of the coordinate operator
$X_\mu$, i.e.:
\be
X_\mu \rket{\xi} = \xi_\mu \rket{\xi} \; ,
\ee
while $P_\mu $ is the momentum operator which satisfies
\bea
\left[ P_\mu , X_\nu \right]_- &=& {\rm i} g_{\mu \nu}  \; ,
\nn \\
\left[ P_\mu , P_\nu \right]_- &=& {\rm i} g G_{\mu \nu}   \; .
\eea
In the coordinate basis $P_\mu$ acts as the covariante derivative
\be \label{eq11a}
\rbra{\xi} P_\mu \rket{\xi'} = {\rm i} \frac{\partial }{\partial \xi^\mu}
\delta (\xi - \xi') + g A_\mu (\xi) \delta (\xi - \xi') \; .
\ee
In this representation the propagator for massless quarks can be
written as
\be
S(\xi ,0) = \rbra{\xi} \frac{1}{\Pslash} \rket{0} \; , \enspace
S(0 ,\xi ) = - \rbra{0} \frac{1}{\Pslash} \rket{\xi} \; .
\ee
In addition to derive equation (\ref{eq7}) we have used the following
identities:
\be
e^{-{\rm i} q \cdot \xi} P_\mu = (P_\mu + q_\mu ) e^{-{\rm i} q \cdot \xi}
\; , \enspace e^{-{\rm i} q \cdot \xi} \rket{0} = \rket{0} \; .
\ee
To avoid misunderstanding in the following we will denote by round
brackets, e.g. $\rbra{0} \dots \rket{\xi}$, the matrix elements in
the formal coordinate space, to be distinguished from square brackets
denoting matrix elements between physical hadronic states, e.g.
$\bra{PS} \dots \ket{PS}$.
The Dirac spinor of the quark of flavor $f$ is denoted  $\psi_f$ and
$q_f$ is its charge.
The sum runs over the light flavors (u,d,s)
in the nucleon under consideration.
Note that $P$ denotes the impuls \it operator \rm in these formulas, while
$q$ and $\xi$ are $c$--numbers.
The matrix element between proton states in (\ref{eq7}) is taken at
the scale $Q^2$.
To evaluate equation (\ref{eq7}) we expand $1/(\Pslash +
\qslash )$ and $1/(\Pslash - \qslash )$:
\bea  \label{eq8}
\frac{1}{\Pslash + \qslash} &=& \frac{1}{\qslash} - \frac{1}{\qslash}
\Pslash \frac{1}{\qslash}
+ \frac{1}{\qslash} \Pslash \frac{1}{\qslash} \Pslash \frac{1}{\qslash}
- \frac{1}{\qslash} \Pslash \frac{1}{\qslash} \Pslash \frac{1}{\qslash}
  \Pslash \frac{1}{\qslash} + \dots  \; ,
\nn \\
\frac{1}{\Pslash - \qslash} &=& - \frac{1}{\qslash} - \frac{1}{\qslash}
\Pslash \frac{1}{\qslash}
- \frac{1}{\qslash} \Pslash \frac{1}{\qslash} \Pslash \frac{1}{\qslash}
- \frac{1}{\qslash} \Pslash \frac{1}{\qslash} \Pslash \frac{1}{\qslash}
  \Pslash \frac{1}{\qslash} + \dots \; .
\eea
The l.h.s. is some
non--local operator, while the r.h.s. gives local differential
operators by identifying $P_\mu = {\rm i} D_\mu = {\rm i}
\partial_\mu + g A_\mu $, with the covariant
derivative  $D_\mu$ (see Eq. (\ref{eq11a})).
Inserting only the first term of the expansion
into equation (\ref{eq7}) yields $T^A_{\mu \nu}$ in the lowest
order.
\bea \label{eq10}
T^A_{\mu \nu} &=& - \frac{1}{q^2} \sum_f \int d^4 \xi \bra{PS} \rbra{\xi}
\bar{\psi}_f q^2_f \left( \gamma_\mu \qslash \gamma_\nu -
\gamma_\nu \qslash \gamma_\mu \right) \psi_f \rket{0} \ket{PS}
\nn \\
&=& 2 {\rm i} \varepsilon_{\mu \nu \lambda \sigma }
\frac{q^\lambda}{q^2} \sum_f \bra{PS} \bar{\psi}_f (0) q^2_f
\gamma^\sigma \gamma^5 \psi_f (0) \ket{PS}
\nn \\
&=& {\rm i} \varepsilon_{\mu \nu \sigma \lambda }
\frac{q^\lambda S^\sigma}{\nu} \;
\omega \; ( 2 \sum_f q_f^2 a_f^{(0)} ) \; ,
\eea
where we have used the identity
\be
\gamma_\mu \gamma_\lambda \gamma_\nu -
\gamma_\nu \gamma_\lambda \gamma_\mu =
2 {\rm i} \varepsilon_{\mu \nu \lambda \sigma} \gamma^5 \gamma^\sigma
\; ,
\ee
($ \gamma^5 = {\rm i} \gamma^0 \gamma^1
\gamma^2 \gamma^3 $)
and have introduced the matrix element of the local operator
\be
\bra{PS} \bar{\psi}_f (0)  \gamma^\sigma \gamma^5
\psi_f (0) \ket{PS} = 2 S^\sigma a_f^{(0)} \; .
\ee
For $T^A_{\mu \nu}$ in the lowest order
this gives:
\bea \label{eq13}
\int_0^1 \left( g_1 (x) + g_2 (x) \right)  \; dx &=& \frac{1}{2}
a^{(0)}
\nn \\
\int_0^1 g_2 (x) \; dx &=& 0  \; ,
\eea
with $a^{(0)} = \sum_f q^2_f a_f^{(0)}$.
This is the well known result:
\be \label{eq14}
\int_0^1 g_1 (x) \; dx = \frac{1}{2} a^{(0)} \; .
\ee

The corrections to this sum rule are obtained by calculating the
next term contributing to $T^A_{\mu \nu}$ denoted by
$\delta T^A_{\mu \nu}$.
\be \label{eq15}
\delta T_{\mu \nu}^A = - \frac{1}{q^6}
\sum_f
\int d^4 \xi \bra{PS} \rbra{\xi}
\bar{\psi}_f q^2_f \left( \gamma_\mu \qslash \Pslash
\qslash \Pslash \qslash \gamma_\nu -
\gamma_\nu \qslash \Pslash \qslash \Pslash \qslash
\gamma_\mu \right) \psi_f \rket{0} \ket{PS} \; .
\ee
Using the equations of motion we get\footnote{
due to the EOM $\int d^4 \xi
\rbra{\xi} \bar{\psi}_f \dots \Pslash \psi_f \rket{0} = 0 =
\int d^4 \xi
\rbra{\xi} \bar{\psi}_f \Pslash \dots \psi_f \rket{0} $}:
\be \label{eq16}
\gamma_\mu \qslash \Pslash \qslash \Pslash \qslash \gamma_\nu -
\gamma_\nu \qslash \Pslash \qslash \Pslash \qslash \gamma_\mu
=
4 \varepsilon_{\mu \nu \lambda \delta} \varepsilon^\delta_{\enspace \alpha
\beta \sigma} q^\lambda q^2 \enspace P^\alpha P^\beta \gamma^\sigma
- 8 {\rm i} \varepsilon_{\mu \nu \lambda \sigma} q^\lambda q_\alpha q_\beta
\enspace P^\alpha P^\beta \gamma^\sigma \gamma^5 \; .
\ee
In the above formula we suppressed $\int d^4 \xi
\rbra{\xi} \bar{\psi}_f \dots \psi_f \rket{0}$.
Now we decompose the Lorentz tensor of rank three $P^\alpha P^\beta
\gamma^\sigma \gamma^5 $ into parts with spin three to zero:
\bea \label{eq17}
&&P^\alpha P^\beta \gamma^\sigma \gamma^5 =
\nn \\ &&
O_{5 \enspace {\rm spin} \; 3}^{\alpha \beta \sigma } +
\tilde{O}_{5 \enspace {\rm spin} \; 2}^{\alpha \beta \sigma}
+ \frac{5}{18} g^{\alpha \beta} P^2 \gamma^\sigma \gamma^5
- \frac{1}{18} g^{\alpha \sigma} P^2 \gamma^\beta \gamma^5
- \frac{1}{18} g^{\beta  \sigma} P^2 \gamma^\alpha \gamma^5
\nn \\
&&
+ \frac{{\rm i} }{6} \varepsilon^{\alpha \beta \sigma \lambda}
  g \tilde{G}_{\lambda \delta } \gamma^\delta \gamma^5 \; ,
\eea
and equivalent for $P^\alpha P^\beta \gamma^\sigma$.
$\tilde{G}_{\lambda \delta }$ is the dual QCD field tensor
\be
\tilde{G}_{\lambda \delta } = \frac{1}{2} \varepsilon_{\lambda \delta
\alpha \beta} G^{\alpha \beta} \; .
\ee
The operators $O$ are defined as follows:
\bea \label{eq19}
&&O_{5 \enspace {\rm spin} \; 3}^{\alpha \beta \sigma }
=
\frac{1}{6} \left(
  P^\alpha  P^\beta  \gamma^\sigma
+ P^\alpha  P^\sigma \gamma^\beta
+ P^\beta   P^\alpha \gamma^\sigma \right.
\nn \\ &&
\left.
+ P^\sigma  P^\alpha \gamma^\beta
+ P^\beta   P^\sigma \gamma^\alpha
+ P^\sigma  P^\beta  \gamma^\alpha
\right) \gamma^5
\nn \\ &&
- \frac{1}{18} \left(
P^2 g^{\alpha \beta} \gamma^\sigma
+ P^2 g^{ \beta \sigma} \gamma^\alpha
+ P^2 g^{\alpha \sigma} \gamma^\beta \right) \gamma^5  \; ,
\\
\label{eq20}
\tilde{O}_{5 \enspace {\rm spin} \; 2}^{\alpha \beta \sigma } &=&
\frac{1}{3} \left(
2 P^\alpha  P^\beta  \gamma^\sigma
- P^\sigma  P^\alpha \gamma^\beta
- P^\beta   P^\sigma \gamma^\alpha
\right) \gamma^5
\nn \\ &&
- \frac{1}{9} \left(
2 P^2 g^{\alpha \beta} \gamma^\sigma
- P^2 g^{ \beta \sigma} \gamma^\alpha
- P^2 g^{\alpha \sigma} \gamma^\beta \right) \gamma^5 \; .
\eea
The operators $O_{ {\rm spin} \; 3}^{\alpha \beta \sigma }$ and
$\tilde{O}_{ {\rm spin} \; 2}^{\alpha \beta \sigma } $
are the same as $O_{5 \enspace {\rm spin} \; 3}^{\alpha \beta \sigma
}$ and $\tilde{O}_{5 \enspace {\rm spin} \; 2}^{\alpha \beta \sigma }$, they
just go without the $\gamma^5$ matrix.
The spin one part of equation (\ref{eq17}) is checked to be
the correct one by taking suitable traces,
the spin zero part by contraction with $\varepsilon_{\alpha \beta
\sigma \lambda'}$.
The spin three and two operators are traceless.
The spin three part is known to have the form of equation
(\ref{eq19}) \cite{MoHa62}.
$\tilde{O}_{5 \enspace {\rm spin} \; 2}^{\alpha \beta \sigma }$
is now determined unique, because it is the only operator that
fulfills the identity (\ref{eq17}).
Although this spin two operator has no specific symmetry properties,
it can be written as a linear combination of two tensors with mixed symmetry.
This corresponds to the fact, that a tensor of rank three contains two
irreducible representations with spin two (as three vectors can be
coupled to spin two in two different ways).

Inserting equations (\ref{eq16}) and (\ref{eq17}) into equation
(\ref{eq15}) gives:
\bea \label{eq21}
&& \delta T_{\mu \nu}^A
=
\nn \\ &&
= \sum_f \left[
- \bra{PS} \bar{\psi}_f (0) q^2_f \left(
O_{ {\rm spin} \; 3}^{\alpha \beta \sigma } +
\tilde{O}_{ {\rm spin} \; 2}^{\alpha \beta \sigma}
+ \frac{5}{18} g^{\alpha \beta} P^2  \gamma^\sigma
- \frac{1}{18} g^{\alpha \sigma} P^2  \gamma^\beta
  \right.  \right.
\nn  \\ &&
\left.
- \frac{1}{18} g^{\beta  \sigma} P^2  \gamma^\alpha
+ \frac{{\rm i} }{6} \varepsilon^{\alpha \beta \sigma \rho}
  g \tilde{G}_{\rho \delta }  \gamma^\delta
\right)
\psi_f (0) \ket{PS}
\cdot
4 \varepsilon_{\mu \nu \lambda \delta'}
\varepsilon^{\delta'}_{\enspace \alpha
\beta \sigma} q^\lambda \frac{1}{q^4}
\nn \\
&& +
\bra{PS} \bar{\psi}_f (0) q^2_f \left(
O_{5 \enspace {\rm spin} \; 3}^{\alpha \beta \sigma } +
\tilde{O}_{5 \enspace {\rm spin} \; 2}^{\alpha \beta \sigma}
+ \frac{5}{18} g^{\alpha \beta} P^2 \gamma^\sigma \gamma^5
- \frac{1}{18} g^{\alpha \sigma} P^2 \gamma^\beta \gamma^5
\right.
\nn \\ &&
\left. \left.
- \frac{1}{18} g^{\beta  \sigma} P^2 \gamma^\alpha \gamma^5
+ \frac{{\rm i} }{6} \varepsilon^{\alpha \beta \sigma \lambda}
  g \tilde{G}_{\lambda \delta } \gamma^\delta \gamma^5
\right)
\psi_f (0) \ket{PS}
\cdot
8 {\rm i} \varepsilon_{\mu \nu \lambda \sigma} q^\lambda q_\alpha q_\beta
\frac{1}{q^6} \right] \; .
\eea
For symmetry reasons only the following terms contribute to
$\delta T_{\mu \nu}^A$
\bea \label{eq22}
&& \delta T_{\mu \nu}^A
=
\nn \\ &&
= \sum_f \left[ - 4
\bra{PS} \bar{\psi}_f (0) q^2_f
 \frac{{\rm i} }{6} \varepsilon^{\alpha \beta \sigma \rho}
  g \tilde{G}_{\rho \delta }  \gamma^\delta
\psi_f (0) \ket{PS}
\cdot
 \varepsilon_{\mu \nu \lambda \delta'}
\varepsilon^{\delta'}_{\enspace \alpha
\beta \sigma} q^\lambda \frac{1}{q^4}  \right.
\nn \\
&& +
\bra{PS} \bar{\psi}_f (0) q^2_f \left(
O_{5 \enspace {\rm spin} \; 3}^{\alpha \beta \sigma } +
O_{5 \enspace {\rm spin} \; 2}^{\alpha \beta \sigma}
+ \frac{5}{18} g^{\alpha \beta} P^2 \gamma^\sigma \gamma^5
- \frac{1}{18} g^{\alpha \sigma} P^2 \gamma^\beta \gamma^5 \right.
\nn \\ &&
\left.
- \frac{1}{18} g^{\beta  \sigma} P^2 \gamma^\alpha \gamma^5
\right)
\left.
\psi_f (0) \ket{PS}
\cdot
8 {\rm i} \varepsilon_{\mu \nu \lambda \sigma} q^\lambda q_\alpha q_\beta
\frac{1}{q^6} \right] \; .
\eea
Note that due to the contraction only $O_5$, but not $O$ contribute
to $g_1$ and $g_2$.
Additional only that part of
$\tilde{O}_{5 \enspace {\rm spin} \; 2}^{\alpha \beta \sigma}$ that is
symmetric in $\alpha \beta$
($O_{5 \enspace {\rm spin} \; 2}^{\alpha \beta \sigma}$)
contributes.
With
\bea
\tilde{O}_{5 \enspace {\rm spin} \; 2}^{\alpha \beta \sigma}
&=&
 \frac{1}{2}
\left( ({\bf 1} + {\bf P}_{\alpha \beta})
\tilde{O}_{5 \enspace {\rm spin} \; 2}^{\alpha \beta \sigma}
+
({\bf 1} - {\bf P}_{\alpha \beta})
\tilde{O}_{5 \enspace {\rm spin} \; 2}^{\alpha \beta \sigma}
\right)
\nn \\
&=&  O_{5 \enspace {\rm spin} \; 2}^{\alpha \beta \sigma}
+ \frac{1}{2}
\left(
({\bf 1} - {\bf P}_{\alpha \beta})
\tilde{O}_{5 \enspace {\rm spin} \; 2}^{\alpha \beta \sigma}
\right) \; ,
\eea
we get
\bea
&&O_{5 \enspace {\rm spin} \; 2}^{\alpha \beta \sigma } =
\nn \\ &&
\frac{1}{6} \left(
2 P^\alpha  P^\beta  \gamma^\sigma
+ 2 P^\beta  P^\alpha  \gamma^\sigma
- P^\sigma  P^\alpha \gamma^\beta
- P^\sigma  P^\beta \gamma^\alpha
- P^\beta   P^\sigma \gamma^\alpha
- P^\alpha   P^\sigma \gamma^\beta
\right) \gamma^5
\nn \\ &&
- \frac{1}{9} \left(
2 P^2 g^{\alpha \beta} \gamma^\sigma
- P^2 g^{ \beta \sigma} \gamma^\alpha
- P^2 g^{\alpha \sigma} \gamma^\beta \right) \gamma^5   \; .
\eea
${\bf P}_{\alpha \beta}$ is the operator, which changes the indices
$\alpha$ and $\beta$.
As stressed by Shuryak and Vainshtein \cite{ESh82}, every step of
this calculation is gauge invariant
($0 = q^\mu T_{\mu \nu} = q^\nu T_{\mu \nu}$).

The matrix elements of the relevant local operators are as follows:
\bea \label{eq27}
&&
\bra{PS} \bar{\psi}_f (0)
O_{5 \enspace {\rm spin} \; 3}^{\alpha \beta \sigma }
\psi_f (0) \ket{PS} =
\nn \\
&&
=  2 a_f^{(2)} \left[ \frac{1}{6}
\left( P^\alpha P^\beta S^\sigma
       + P^\beta  P^\alpha S^\sigma
       + P^\sigma  P^\alpha S^\beta
       + P^\beta  P^\sigma S^\alpha
       + P^\alpha  P^\sigma S^\beta
       + P^\sigma P^\beta S^\alpha
\right)  \right.
\nn \\
&&
- \frac{M^2}{18}
\left. \left( g^{\alpha \beta} S^\sigma + g^{\beta \sigma} S^\alpha +
       g^{\alpha \sigma} S^\beta
\right) \right]  \; ,
\eea

\bea \label{eq28}
&&
\bra{PS} \bar{\psi}_f (0)
O_{5 \enspace {\rm spin} \; 2}^{\alpha \beta \sigma }
\psi_f (0) \ket{PS} =
\nn \\ &&
=  \frac{1}{6} \bra{PS} \bar{\psi}_f (0) \left[ \gamma^\alpha g
\tilde{G}^{\beta \sigma } + \gamma^\beta g \tilde{G}^{\alpha \sigma }
\right] \psi_f (0) \ket{PS} - {\rm traces } =
\nn \\ &&
=
  2 d_f^{(2)} \left[ \frac{1}{6}
\left(
2 P^\alpha P^\beta S^\sigma + 2 P^\beta P^\alpha S^\sigma
- P^\beta P^\sigma S^\alpha - P^\alpha P^\sigma S^\beta
- P^\sigma P^\alpha S^\beta - P^\sigma P^\beta S^\alpha
\right) \right.
\nn \\
&&
\enspace - \frac{M^2}{9}
\left. \left(
2 g^{\alpha \beta} S^\sigma - g^{\beta \sigma} S^\alpha - g^{\alpha
\sigma} S^\beta
\right) \right]   \; ,
\eea

\be \label{eq29}
\bra{PS} \bar{\psi}_f (0)
P^2 \gamma_\alpha \gamma_5  \psi_f (0)
\ket{PS} =
\bra{PS} \bar{\psi}_f (0)
g \tilde{G}_{\alpha \beta} \gamma^\beta
\psi_f (0) \ket{PS} =
2 M^2 f_f^{(2)} S_\alpha \; .
\ee
With this definitions
\footnote{Note the sign difference in equation (\ref{eq29}) with
respect to
Ji and Unrau \cite{XJi93}, owing to the fact, that they use
$D_\mu = \partial_\mu + {\rm i} g A_\mu $. }
for the matrix elements and equation
(\ref{eq12}) one gets from $T_{\mu \nu}^A$ and
$\delta T_{\mu \nu}^A$:
\bea \label{eq30}
\int_0^1 dx \; \left( g_1 (x,Q^2) + g_2 (x,Q^2) \right)
&=&
\frac{1}{2} a^{(0)} + \frac{M^2}{9 Q^2}
\left( a^{(2)} + 4 d^{(2)} + 4 f^{(2)} \right) \; ,
\nn \\
\int_0^1 dx \;  g_2 (x,Q^2)
&=&
0 \; ,
\nn \\
\int_0^1 dx \; \left( g_1 (x,Q^2) + g_2 (x,Q^2) \right) x^2
&=&
\frac{1}{6} a^{(2)} + \frac{1}{3} d^{(2)} \; ,
\nn \\
\int_0^1 dx \; g_2 (x,Q^2) x^2
&=&
- \frac{1}{3} a^{(2)} + \frac{1}{3} d^{(2)} \; ,
\eea
with
\be \label{eq35}
a^{(n)} = \sum_f q^2_f a_f^{(n)} \; , \enspace
d^{(n)} = \sum_f q^2_f d_f^{(n)} \; , \enspace
f^{(n)} = \sum_f q^2_f f_f^{(n)} \; .
\ee
Including the next term of the expansion in $T_{\mu \nu}$ would
result in corrections of the order $(Q^2/M^2)^2$ for the first
moments, corrections of the order $Q^2/M^2$ for the third moments,
and it would give values for the fifth moments.
We can therefore conclude:
\bea \label{eq31}
\int_0^1 dx \; g_1 (x,Q^2)
&=&
\frac{1}{2} a^{(0)} + \frac{M^2}{9 Q^2}
\left( a^{(2)} + 4 d^{(2)} + 4 f^{(2)} \right) + O ( \frac{M^4}{Q^4} )
\; ,
\\ \label{eq32}
\int_0^1 dx \;  g_2 (x,Q^2)
&=&
0 + O ( \frac{M^4}{Q^4} ) \; ,
\\ \label{eq33}
\int_0^1 dx \; g_1 (x,Q^2)  x^2
&=&
\frac{1}{2} a^{(2)} + O ( \frac{M^2}{Q^2} )  \; ,
\\ \label{eq34}
\int_0^1 dx \; g_2 (x,Q^2) x^2
&=&
- \frac{1}{3} a^{(2)} + \frac{1}{3} d^{(2)} + O ( \frac{M^2}{Q^2} )
\; .
\eea
Equation (\ref{eq32}) states that
up to twist four and in the leading order of $\alpha_s$
there are no terms which
contribute to the first moment of $g_2$, i.e. which would violate the
Burkhardt--Cottingham sum rule \cite{HBu70}.
The sign in front of $f^{(2)}$ in equation (\ref{eq31}) is positive.
This is in agreement with the result of Shuryak and Vainshtein
\cite{ESh82}
and Balitsky, Braun, Kolesnichenko \cite{IIBa90}.
In their revised paper \cite{XJi93} Ji and Unrau have derived
the $f^{(2)}$ term with a negative sign, but this is only due to their
convention for $D_\mu$, see footnote to equation (\ref{eq29}).
We want to stress that all three calculations agree with each other
when the differences due to different conventions are taken into
account.

With $a^{(2)}_{{\rm proton}} = 0.022 \pm 0.002$ from the E130 \cite{E130}
and the EMC \cite{EMC} measurement,
$a^{(0)}_{{\rm NS}} $ and $a^{(0)}_{{\rm S}} $ from the EMC
measurement,
and the estimates
$d^{(2)}_{{\rm proton}} = - 1.4 \cdot 10^{-3} \pm  4.0 \cdot 10^{-3}$
($d^{(2)}_N = - 0.026 \pm  0.004 $) and $f^{(2)}_{{\rm proton}} =
- 0.050 \pm 0.018$ ($f^{(2)}_N =  0.018 \pm 0.018$),
which we get from the QCD sum rule calculation of
Balitsky, Braun and Kolesnichenko \cite{IIBa90}
one obtaines
\bea \label{eq38}
&&\int_0^1 g_1^p (x,Q^2) \; dx =
\\
&& =
  0.124 \cdot (1 - \frac{\alpha_s (Q^2)}{\pi} )
+ 0.013 \cdot (1 - \frac{33 - 8 n_f}{33 - 2 n_f} \frac{\alpha_s (Q^2)}{\pi} )
- (0.018 \pm 0.0072 ) \cdot \frac{{\rm GeV}^2}{Q^2}  \; . \nn
\eea
The QCD radiative corrections are taken from \cite{JKo79}.
Figure 2 shows this dependence of the first moment of $g_1$ on $Q^2$
for low $Q^2$ ($n_f = 3$).
Our result agrees with the result of ref. \cite{IIBa90}.
Note, however, that in their publications $\gamma^5_{BBK} = - \gamma^5$ and
therefore $S_{\sigma \enspace BBK} = \bar{N} \gamma_\sigma \gamma^5_{BBK}
N = - 2 S_\sigma$.
The different sign in front of the $Q^2$ corrections of equation
(\ref{eq38}) compared to the sign of Ji and Unrau \cite{XJi93}
(equation (29)) is due to fact,
that although their calculation agrees with ours, numerically the bag
model prediction for the matrix element $f^{(2)}_{{\rm proton}}$
has opposite sign than the predictions from QCD sum rules.

This work was supported in part by DFG (G. Hess Program).
\enspace A.S. thanks also the MPI f\"ur Kernphysik in Heidelberg for its
hospitality.
L.M. was supported in part by A. v. Humboldt foundation and in part
by KBN under grant 2~P302~143~06.
The authors thank V.M.~Braun X.~Ji and E.~Stein for helpful conversation.

\newpage

{\LARGE \bf Figure Caption}
\rm
\\ \\ \\
Figure 1:
Graphs contributing to $T \left( j_\mu (\xi) \; j_\nu (0) \right)$.

Figure 2:
Dependence of the first moment of $g_1^P$ on $Q^2$ (solid line).
The dotted line shows only the radiative corrections.
The dashed line corresponds to $- 0.018/Q^2$.

\end{document}